\documentclass[aps,pra,showpacs,eqsecnum,twocolumn]{revtex4}
\usepackage{graphicx}
\usepackage{bm}
\usepackage{amsfonts}

\usepackage{amssymb,amsthm,dsfont,bm,color,ulem}

\begin{document}
\preprint{}
\title{Nonlinear fiber gyroscope for  quantum metrology}
\author{Alfredo Luis}
\email{alluis@fis.ucm.es}
\homepage{http://www.ucm.es/info/gioq}
\author{Irene Morales}
\affiliation{Departamento de \'{O}ptica, Facultad de Ciencias
F\'{\i}sicas, Universidad Complutense, 28040 Madrid, Spain}
\author{\'{A}ngel Rivas}
\affiliation{Departamento de F\'isica Te\'orica I, Facultad de Ciencias F\'isicas,
Universidad Complutense, 28040 Madrid, Spain}

\date{\today}

\begin{abstract}
We examine the performance of a nonlinear fiber gyroscope for improved signal detection
beating the quantum limits of its linear counterparts. The performance is examined when
the nonlinear gyroscope is illuminated by  practical field states,  such as coherent and
quadrature squeezed states. This is compared with the case of more ideal probes such
as photon-number states.
\end{abstract}


\maketitle

\section{Introduction}

Signal-detection strategies based on nonlinear processes can clearly outperform current
strategies based on linear processes
\cite{LU04,BFCG07,BDDFSC08, BDDSTC09,GZNEO10,NM10,TBDSC10,RL10,LU10,NKDBSM11,SNBCCM14,LR15,JLZLBBNDW13}.
This is so even when using probes in classical-like states. This is relevant because classical-like states
are  characterized by their robustness against practical imperfections, that can be deadly for
schemes using probes prepared in nonclassical states \cite{HMPEPC97,DBS13,DJK14}.

A suitable arena for nonlinear detection schemes is optics. The most precise detection schemes
are optical interferometers, and nonlinear processes are quite simply implemented in optics via
propagation in nonlinear media.

In this work we focus on the quantum limits to the resolution achievable in gyroscopes as
good candidates to exploit the benefits of nonlinear detection, paying special attention
to quantum characteristics exclusive of this interferometer   \cite{GY}. We will focus on the case when the
gyroscope is illuminated by  practical field states, i. e., that can be generated in practice and are
robust against imperfections, such as coherent and quadrature squeezed states. For the sake of
comparison the results will be compared with the case of probes in less practical
states, such as product of number states.

There are two important advantages in the proposed scheme with respect to other nonlinear
detection schemes, namely, its improved optical performance and its capability to measure angles.
More specifically, as a comparison with other interferometers (like Michelson's) nonlinearity is
integrated as a constituent part of the gyroscope, so that it is not necessary to modify the set up to
include extra elements to provide the nonlinear effect. For instance,  in order to include nonlinearity
in the LIGO gravitational-wave detector  we should ``attach'' some nonlinear material to it  \cite{LIGO}.
The simplest way to do this would be by filling the room with a nonlinear gas.
This will produce  severe practical inconvenient since gases are a reported source of  technical noise
because of density fluctuations and light scattering, which is the reason why LIGO works at ultra-hight
vacuum. Furthermore the nonlinearity in gases is typically small. An alternative would be to attach
a piece of nonlinear material to the suspended masses. In such a case some light will be reflected
in the interphase and in any case it far for technically simple to make such attachments because of the
large dimensions of the device. Our proposal is free of these problems because the nonlinearity is
automatically embedded as a part of the interferometer: the optical fiber. The optical performance
of fiber glasses is much superior than gases regarding homogeneities and any other source of
scattering and fluctuations. Moreover, the nonlinear effect is much larger than for gases, being
improved by the very large electric fields that can be reached by light confinement in the small
volumes of the fiber core. Gyroscopes do not need to be kilometer long devices to provide extremely
long optical paths required for precision interferometry since looped fiber-optics coil multiplies the
length and the  cumulative effects of nonlinearlity by the number of loops. Finally, the gyroscope is
a rigid detector without moving parts, that  always provides a more robust and improved optical
performance.

On the other hand, up to our knowledge nobody has considered to apply nonlinear quantum
metrology  in a gyroscopic scheme. This is very timely because there are many interesting
unobserved effects caused by rotations, specially to test gravitational theories and phenomena.
For example, the Lense-Thirring effect, a relativistic effect not observed yet, the local space-time
curvature, or the existence of a preferred frame in the Universe \cite{OZ}.

Besides fiber-optics realizations, previous works have shown that nonlinear interferometers are feasible
in other physical contexts, such as Bose-Einstein  condensates \cite{BDDFSC08,BDDSTC09,GZNEO10,TBDSC10,JD}
and nanomechanical resonators \cite{WMC}. Finally, we may point out that  gyroscopes share geometry
with sensors sensible to physical variables different from rotations and involving relevant physical phenomena
such as the Aharonov-Bohm effect for example \cite{GDR}.

\section{Model}

Any detection scheme involves four steps. In the first one, some probe state  $| \psi \rangle$ is
prepared. In the second step  the probe experiences a signal-dependent transformation $U (\phi)$.
Then, the measurement of some observable $M$ is performed at the output state of the probe
$U (\phi) | \psi \rangle$. With the results of the measurement the signal $\phi$ and its uncertainty
$\Delta \phi$ are estimated.

\subsection{Probe and system modes}
The probe  $| \psi \rangle$ is the light state illuminating the gyroscope.  Within the interferometer
the system is made of two counter-propagating modes with complex-amplitude operators
$a_\pm$, that can be feed  trough a lossless 50 \%  beam splitter coupling the inner modes $a_\pm$
with two input modes $a_{1,2}$ (see Fig. 1)
\begin{equation}
\label{iom}
a_\pm = \frac{1}{\sqrt{2}} \left ( a_1 \mp i a_2 \right )  .
\end{equation}
The probes are prepared in modes $a_{1,2}$ while the observation will be made at the output
modes $\bar{a}_{1,2}$  leaving the gyroscope. These are related to the modes within the
interferometer through a beam splitter performing the transformation inverse to the one in Eq. (\ref{iom})
\begin{equation}
\label{iom2}
\bar{a}_1 = \frac{1}{\sqrt{2}} \left ( a^\prime_+ + a^\prime_- \right ), \quad
\bar{a}_2 = \frac{i}{\sqrt{2}} \left ( a^\prime_+ - a^\prime_- \right ) ,
\end{equation}
where $a^\prime_\pm$ are the amplitudes at the end of the nonlinear fiber, while
$a_\pm$ refer to the amplitudes at the beginning.

\begin{figure}
\begin{center}
\includegraphics[width=7cm]{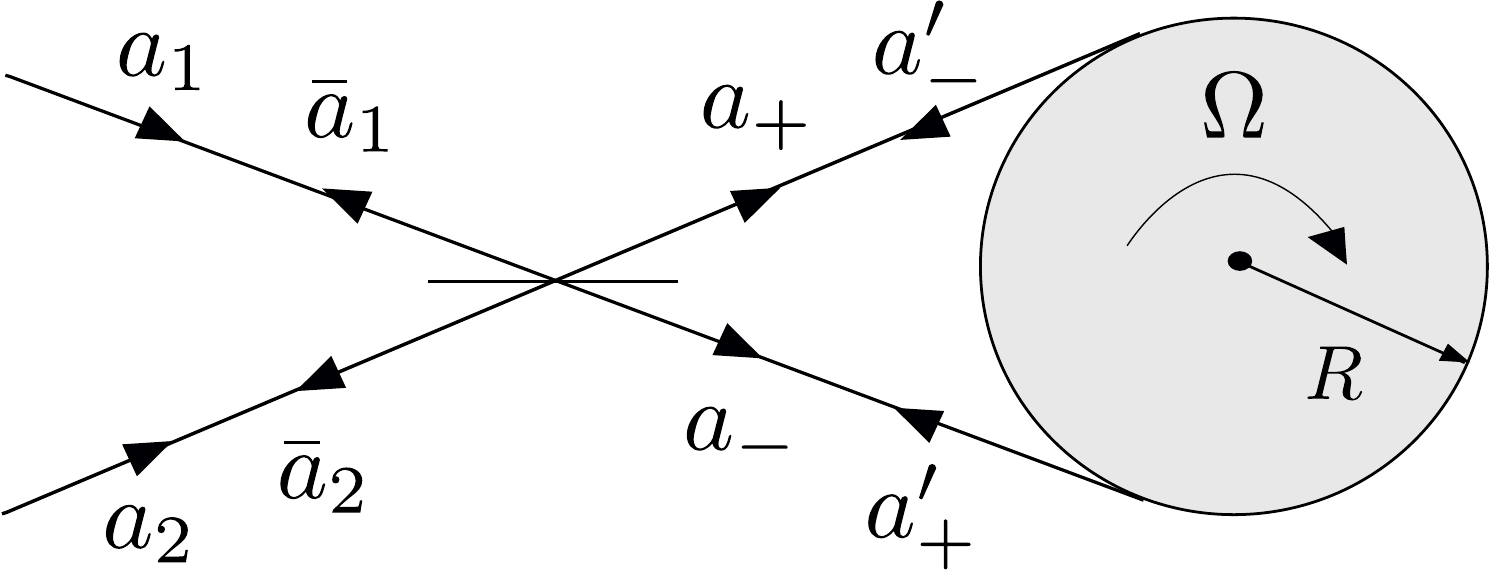}
\end{center}
\caption{Scheme of the gyroscope illustrating the definition of the field modes.}
\end{figure}

\subsection{Signal-dependent transformation}
In the second step  the probe experiences a signal-dependent transformation $U (\phi)$.
The rotation of the gyroscope introduces an asymmetry between the times spent
 by the two modes  $a_\pm$  within the interferometer, that leads to a phase-difference
 \begin{equation}
 \label{pd}
 \varphi \simeq \frac{\omega}{c} L \left ( n_+ - n_- \right ) + 2 \frac{\omega A  \mathcal{N}}{c^2} \Omega
 \left ( n^2_+ + n^2_-\right ) ,
 \end{equation}
where $\omega$ is the field frequency, $L$ is the length of the fiber, $A = \pi R^2$ is the area
enclosed by  a single loop of the fiber made of loops of radius $R$, $\mathcal{N}$ is the number of
loops,  $\Omega$ is the angular speed,  $n_\pm$ are the indices of
refraction for the corresponding modes, and we have assumed that the speed acquired
by the fiber due to $\Omega$ is much smaller than the speed of light in vacuum $c$.

The key point for our work is that  we are dealing with nonlinear media so the indices of
refraction $n_\pm$ depend on the light intensities. Assuming Kerr-type nonlinearity and
light traveling as pulses of frequency $\omega$, cross-section $\cal{A}$, and duration
$\tau$, carrying a number of photons $N_\pm$, we will have
\begin{equation}
\label{in}
n^2_\pm \simeq n_0^2 + \chi_\pm E_\pm^2 \simeq n_0^2 + \frac{\mu_0 \hbar \omega c
\chi_\pm}{\cal{A} \tau} N_\pm
\end{equation}
where  $n_0$ is the linear index, $\chi_\pm$ are the nonlinear susceptibilities, for simplicity
the medium is assumed optically isotropic so that the linear index is the same for both modes,
$E_\pm$ are the electric-field strengths, and $\mu_0$ is the magnetic permeability of the vacuum. The
second equality in Eq. (\ref{in}) is just a rough approximation to motivate the ongoing quantum
analysis. We have assumed that there are no crossed terms. This is specially so in a
pulsed illumination since in such a case the overlap of the counter propagating pulses is negligible.

In the quantum analysis, the propagation of the modes $a_\pm$ within the fiber can be described
by the unitary operator $U_\Omega U_0$, where $U_\Omega$ includes all the signal-dependent
effects given by the second term on the right-hand side of Eq. (\ref{pd}), while $U_0$ contains the
contributions independent of $\Omega$, this is the contribution of the first term on the right-hand
side of Eq. (\ref{pd}). The factorization is possible because both parts can be fully expressed in
terms of two commuting phase shifts generated by different powers of the photon-number operators.

In this work we focus on the nonlinear part of the signal-dependent component  $U_\Omega$.
The contribution by $U_0$ will be invoked just to introduce and additional fixed phase $\phi_0$
when necessary. Otherwise, it will be assumed embodied in the probe preparation or
compensated by a similar amount of fiber propagation not experiencing the rotation.

For the signal-dependent part $U_\Omega$ we will just consider the nonlinear contribution
given by the second term on the right-hand side of Eq. (\ref{in}). This should
be the dominating part for  sufficiently large photon numbers, as far as we intend to exploit
the asymptotic behavior allowed by  nonlinear effects. In any case, the phase shifts
produced by the linear and nonlinear parts of the transformation may be addressed
simultaneously via a multi-parameter estimation procedure \cite{Ch}. This is studied in
more detail in Appendix B showing that it agrees with the expected results both for the
linear and nonlinear parts.

Thus, the signal-dependent transformation we are going to study is
\begin{equation}
\label{UO}
U_\Omega = e^{- i \phi G} , \qquad  G = N_+^2 - N_-^2   ,
\end{equation}
where $N_\pm = a^\dagger_\pm a_\pm $ are the corresponding number operators and
the signal takes the form $\phi = \mu_0 A \hbar \omega^2 \mathcal{N} \chi \Omega /({\cal A} \tau c)$,
assuming the nonlinear susceptibilities identical for both modes $\chi_+ = \chi_- = \chi$.
Note that the relative sign in Eq. (\ref{UO}) is the correct one for counter-propagating modes.
The sing depending on the propagation direction is often expressed by saying that for
counter-propagating modes the generator $G$ is proportional to momentum rather than
to energy \cite{mg}.

Using Eq. (\ref{iom}) we find an useful expression for the generator $G$, in terms of the input
modes $a_{1,2}$ by
\begin{equation}
\label{G}
G = N \left ( N_+  - N_- \right ) =  i N \left ( a^\dagger_2 a_1 - a^\dagger_1 a_2 \right ) ,
\end{equation}
where $N$ is the total-number operator
\begin{equation}
N = N_+  + N_- = N_1 +  N_2 ,
\end{equation}
and we have used that $N$ is a conserved quantity at the beam splitter. We note that the second
equality in (\ref{G})  is formally the same generator already tested experimentally  in
Ref. \cite{NKDBSM11} in a very different context.

\subsection{Measurement}
The most simple measurement sensitive to signals encoded as phase shifts is  the interference
achieved by the coupling at the beam splitter of  the $a_\pm$ modes after the nonlinear
propagation, $a^\prime =  U^\dagger_\Omega a_\pm U_\Omega$, followed by
photon-number detection at the outgoing beams $\bar{a}_{1,2}$. This provides us with
a photon-number statistics $p (n_1, n_2 | \phi)$ containing the complete information about
the signal available in this arrangement. In order to extract this information the most simple
option is to consider the difference between the output photon numbers as $M = \bar{a}_1^\dagger
\bar{a}_1 - \bar{a}_2^\dagger \bar{a}_2$. Since the transformation (\ref{UO}) is quite simple we
have the following expression for $M$ in terms of the complex amplitudes at the beginning
of the fiber $a_\pm$
\begin{equation}
M =  a^\dagger_+ e^{i \Lambda} a_- + a^\dagger_- e^{-i \Lambda} a_+  , \quad
\Lambda = 2 \phi N  + \phi_0,
\end{equation}
and $\phi_0$ is any fixed additional linear phase shift introduced to optimize performance.
Then, we can express $M$ in terms of the input modes $a_{1,2}$ via Eq. (\ref{iom}) as
\begin{equation}
\label{MSC}
M = a^\dagger_1 C a_1- a^\dagger_2 C a_2 - a^\dagger_2 S   a_1 - a^\dagger_1 S a_2   ,
\end{equation}
where $C = \cos \Lambda $, $S = \sin \Lambda$.

For the most simple situations the lowest-order moments of $M$ may be enough. However, there are
situations where they may not extract most of the signal information encoded in the output field
state. In such situations we may look at the complete output photon-number statistics $p \left ( n_1, n_2 |
\phi \right )$ in order to better understand the situation.

\subsection{Simple estimation}
The key performance estimator is the signal uncertainty  $\Delta \phi$.  In a very simple first approach
this can be estimated from the lowest-order moments of $M$ via the signal to noise ratio, or, equivalently,
from a simple error propagation as:
\begin{equation}
\label{Df}
\Delta \phi = \frac{\Delta M}{\left | \partial \langle M \rangle / \partial \phi \right |} =
\frac{\Delta M}{\left | \langle [M , G ] \rangle \right |}  \geq \frac{1}{2 \Delta G} ,
\end{equation}
where in the last inequality the uncertainty relation $\Delta G \Delta M \geq |\langle [G,M] \rangle/2$ has
been used. In some relevant situations optimum results are obtained if $\phi_0 = - \pi/2$ so that for very
small signals $\langle M \rangle$ will be near zero, which is the point of maximum sensitivity to phase
variations. In such a case for  $\phi \rightarrow 0$ we have after Eqs. (\ref{MSC}), (\ref{UO}),  and (\ref{G})
\begin{equation}
\label{M}
M = i \left ( a^\dagger_-  a_+  -  a^\dagger_+ a_-  \right ) =  a^\dagger_1  a_2  +  a^\dagger_2 a_1 ,
\end{equation}
and
\begin{equation}
\label{MG}
 [M , G ] = 2 i N \left ( a^\dagger_+  a_- + a^\dagger_- a_+  \right ) = 2 i \left ( N_1^2 -
 N_2^2  \right ) .
\end{equation}

At this stage one might be tempted to look for optimum results in terms of minimum uncertainty states
of $G$ and $M$, granting the equality in the last step in Eq. (\ref{Df}). However, this is not quite so
optimum strategy since some other probes may lead to smaller $ \Delta \phi$ via a larger $\Delta G$
even though they are not minimum
\cite{DA}.

\subsection{Advanced estimation}
The estimation of the uncertainty $\Delta \phi$ can be addressed using more powerful tools such as the
Cram\'er-Rao lower bound and the quantum Fisher information  $F_Q$ as  \cite{Fi}
\begin{equation}
\label{Df2}
\Delta \phi  \geq \frac{1}{\sqrt{F}}  \geq  \frac{1}{\sqrt{F_Q}}  \geq \frac{1}{2 \Delta G} ,
\end{equation}
where $F$ is the Fisher information
\begin{equation}
F = \sum_{n=0}^\infty \frac{1}{p_n } \left ( \frac{\partial p_n}{\partial \phi} \right )^2 ,
\end{equation}
with $p_n \equiv p \left ( n_1, n_2 | \phi \right )$ and $n$ standing for the pair of natural numbers $(n_1,n_2)$
representing the number of photons registered at the two outputs of the gyroscope. In the case of no additional
phase shift $\phi_0 =0$ the output photon-number statistics is
\begin{eqnarray}
p_n  & = & \left | \langle n_1, n_2 | U^\dagger_{BS} e^{- i \phi G} U_{BS}  | \psi \rangle \right  |^2  \nonumber \\
& = &
\left | \langle n_1, n_2 | e^{ \phi N \left (   a^\dagger_2  a_1 -  a^\dagger_1 a_2  \right )} | \psi \rangle
\right  |^2  ,
\end{eqnarray}
where $U_{BS} $ is the unitary operator representing the action of the input beam splitter, $| n_1, n_2
\rangle$ are number states, and we have taken into account that the output beam splitter performs the
inverse transformation of the input.

In the case that the probe is in a pure state with real coefficients in the number basis $\langle n_1,
n_2 | \psi \rangle \in \mathbb{R}$, then  $p_n = c^2_n$, with  $c_n = \langle n_1, n_2 |  e^{\phi N
\left (   a^\dagger_2  a_1 -  a^\dagger_1 a_2  \right )} | \psi \rangle \in \mathbb{R}$ and it can be easily
shown that \cite{DDLP}
\begin{equation}
F = \sum_{n=0}^\infty 4  \left ( \frac{\partial c_n}{\partial \phi} \right )^2 = 4 \Delta^2 G = F_Q ,
\end{equation}
where in the last step we have used that  if  $\langle n_1, n_2 | \psi \rangle \in \mathbb{R}$ then
$\langle \psi | G | \psi \rangle =0$ after Eq.  (\ref{G}) .

Naturally the fact that the full photon number statistics outperforms those of the simple measurement
of $M$ is rather obvious since the statistics of $M$ is a marginal of the full statistics $p_n$. The key point
here is that when the probe state has real coefficients in the number basis, the $p_n$ statistics contains
all the information conveyed by the transformed probe state so that its Fisher information equals the quantum
Fisher information.
Note that this conclusion holds for all $\phi$. Since the total-number variable $N$ does not provide phase
information, all the phase information in $p_n$ is actually provided by $M$. As we shall see, in many cases
of interest the two lowest-order moments of $M$ already contain all the relevant information about $\phi$.

\section{Quantum resolution for different probes}

Typically, $\Delta \phi$ decreases as the mean number of photons $\bar{N}$  in the probe state $| \psi
\rangle$ increases. So the usual task is to look for the minimum $\Delta \phi$ at fixed  $\bar{N}$, this is
to say, minimum uncertainty at fixed energy resources. Alternatively, this is to inquire about the probe
states that provide the best scaling of $\Delta \phi$ as a function of $\bar{N}$. To this end we will consider
different probe states under the two estimation uncertainties in Secs.  IID and IIE.

\subsection{Product of coherent states}

If the input probe is in a product of Glauber coherent states $|\psi \rangle = | \alpha_1 \rangle | \alpha_2
\rangle$, the field state in modes $a_\pm$ will be as well a product of coherent states $|\psi \rangle =
| \alpha_+ \rangle | \alpha_- \rangle$ with the same total mean number of photons $\bar{N} = | \alpha_+ |^2
+  | \alpha_- |^2  =  | \alpha_1 |^2  +  | \alpha_2 |^2 $. A simple calculus leads exactly to
\begin{equation}
\langle M \rangle = 2 \sqrt{\bar{n}_+ \bar{n}_-} e^{- 2 \bar{N} \sin^2 \phi}
\cos \left [ \bar{N} \sin  \left ( 2 \phi \right ) + \phi_0 \right ] ,
\end{equation}
while
\begin{eqnarray}
& \langle M^2 \rangle = \bar{N}  + 2 \bar{n}_+ \bar{n}_- \left ( 1 + \right . & \nonumber \\
& \left . e^{- 2 \bar{N} \sin^2 (2 \phi )} \cos \left [ \bar{N} \sin  \left ( 4 \phi \right ) + 4 \phi + 2 \phi_0 \right ]  \right ) , &
\end{eqnarray}
where $\bar{n}_\pm = | \alpha_\pm |^2$ and we have assumed real $\alpha_\pm$. Typically $\phi$ is small enough
so that  $ \sin \phi \simeq \phi $ and  $\sin (2 \phi) \simeq 2 \phi$. Moreover, will take advantage of the robustness
of coherent states to consider a large mean number of photons $\bar{N} \gg 1$.

We can observe that the dispersion of total number of photons in the coherent state degrades the
visibility of the interference through the factor $e^{- 2 \bar{N} \phi^2}$. Therefore, in order to obtain
meaningful results  it is convenient to assume that  $\sqrt{\bar{N}} \phi \ll 1$. This is to say that the
signal expected is below the standard quantum limit, so it would pass unnoticed if the fiber were
linear. In such a case,  we have for $\phi_0 = - \pi/2$
\begin{equation}
\langle M \rangle \simeq 2 \sqrt{\bar{n}_+ \bar{n}_-} \sin \left ( 2 \bar{N}  \phi \right ) ,
\end{equation}
and $\Delta^2 M = \bar{N}$ for all $\phi$, so that after Eq. (\ref{Df})
\begin{equation}
\Delta^2 \phi = \frac{1}{16 \bar{N} \bar{n}_+ \bar{n}_- \cos \left ( 2 \bar{N}  \phi \right )  } .
\end{equation}
The optimum result holds for small enough signals $\phi \ll 1/\bar{N}$ well below the Heisenberg
limit of linear devices so that  $\cos \left ( 2 \bar{N}  \phi \right ) \simeq 1$. When varying the balance
of photons between the modes, the minimum uncertainty holds for an equal splitting of resources
between the fiber modes, this is  $\bar{n}_+ = \bar{n}_- = \bar{N}/2$ so that  $|\psi \rangle = | \alpha_1
= \sqrt{\bar{N}} \rangle | \alpha_2 =0 \rangle$ with a coherent state $\alpha = \sqrt{\bar{N}}$ in
mode $a_1$ and vacuum in mode $a_2$. With all this we get
\begin{equation}
\label{Dfv}
\Delta^2 \phi \simeq \frac{1}{4 \bar{N}^3 }.
\end{equation}
Note that is quite below the standard quantum limit and the Heisenberg limit of linear devices, so the
result is consistent with the approximations made.

If we go beyond the simple estimation we have that the probe  $ | \alpha_1 = \sqrt{\bar{N}} \rangle
| \alpha_2 =0 \rangle$  has real coefficients in the number basis. So, if now we consider $\phi_0 =0$,
we get $F = F_Q \simeq  4 \bar{N}^3$  to the leading order in $\bar{N}$ for all $\phi$. As a bonus
we get that the intense coherent states behave as minimum uncertainty states for the $G$, $M$ pair.

\subsection{Product of coherent and squeezed states}

The benefits of using squeezed states in linear quantum metrology are well-known from
a long time ago \cite{sv0,sv1,sv2}. We can check whether a similar result holds in the nonlinear
case. To show this in the simplest manner we consider as probe state  in mode $a_1$ a
coherent state $| \alpha \rangle$ with real $\alpha$ and and squeezed vacuum in mode
$a_2$ with squeezing parameter $r$ and mean number of photons $\bar{N}_2 = \sinh^2 r$.
Again we consider a fixed mean total number of photons $\bar{N} = | \alpha|^2 + \bar{N}_2$
with $\bar{N} \gg 1$.

We begin with the simple evaluation of $\Delta \phi$ in Eq. (\ref{Df}) for $\phi_0 = - \pi/2$ and
$\phi \rightarrow 0$. For the choice of the squeezing direction to reduce fluctuations of the
quadrature $a_2  + a^\dagger_2 $ we have after Eq. (\ref{M})
\begin{equation}
\Delta^2 M =   | \alpha|^2 e^{- 2r} + \bar{N}_2
\end{equation}
and from Eq. (\ref{MG})
\begin{equation}
\langle [M , G ] \rangle = 2 i \left (  | \alpha|^4 +  | \alpha|^2 - \bar{N}_2^2 - \Delta^2 N_2 \right )
\end{equation}
with
\begin{equation}
\Delta^2 N_2 = 2 \sinh^2 r \cosh^2 r .
\end{equation}
For $\alpha \gg1$ and $r \gg 1$ it can be readily seen that the minimum $\Delta \phi$ in Eq. (\ref{Df})
holds for $\bar{N}_2 \simeq \sqrt{\bar{N}}/2 \ll \bar{N}$, leading to
\begin{equation}
\label{sM}
\Delta^2 \phi = \frac{1}{4 \bar{N}^{7/2}}  .
\end{equation}
Comparing Eqs. (\ref{Dfv}) and  Eq. (\ref{sM}) we see that squeezing provides an effective improvement
of resolution over the pure coherent case. Moreover, after noting that in these conditions $\Delta^2 G
\simeq 2 \bar{N}^{7/2}$ we get that the probe is close to be a minimum uncertainty states of the pair $G$,
$M$ since $\Delta^2 G \Delta^2 M \simeq 2 \bar{N}^4$ is twice the minimum $| \langle [M,G]
\rangle |^2 /4 \simeq \bar{N}^4$.

In Appendix A we have carried out an analysis of errors in the presence of several sources of noise that
can be readily accounted for by the replacement in the detection operator $M$ in Eq. (\ref{M})
 \begin{equation}
a^\dagger_1 a_2 \rightarrow e^{i \varphi} \left (\sqrt{\eta} a^\dagger_1 +  \sqrt{1- \eta} \,
b^\dagger_1 \right ) \left (  \sqrt{\eta} a_2 + \sqrt{1- \eta} \, b_2 \right ) ,
 \end{equation}
where $\varphi$ is a random phase Gaussian distributed with zero mean and variance $\sigma^2$, $\eta$
is the quantum efficiency of the detectors, and $b_j$ are uncorrelated field modes in thermal states carrying
$N_t$ photons with $N_t  \ll \bar{N}$. The signal uncertainty becomes
 \begin{equation}
 \Delta^2 \phi \simeq  \frac{1}{4 \eta^2 \bar{N}^{7/2}} + \frac{N_t+1-\eta}{4 \eta^3 \bar{N}^3}  +
 \frac{\sigma^2}{2 \eta^2 \bar{N}^{5/2}}  .
 \end{equation}
The dominant factor are the fluctuations of the relative phase $\varphi$ that  would spoil the improvement
caused by the squeezing vacuum unless $\sigma^2 < 1/\bar{N}$. The other two terms would tend to phase
uncertainty similar to the case of pure coherent probes unless $N_t + 1 - \eta < 1/\sqrt{\bar{N}}$. 

After the tools in Sec. IIE we can go beyond taking into account that when $\alpha$ is real and when the
squeezing takes place either in the quadrature $i( a_2  - a^\dagger_2) $ or in $a_2  + a^\dagger_2 $, the
coefficients of the probe state in the number basis are real so that $F = F_Q$. Then we can ask for the
balance between $\alpha^2$ and $\bar{N}_2$ that leads to maximum $F_Q$ at fixed  $\alpha^2 + \bar{N}_2
= \bar{N}$. For $\alpha \gg 1$ we can neglect the fluctuations in mode $a_1$. Moreover, for  $r \gg1$ we
can carry our the following approximation for the squeezed mode $N_2 \simeq \bar{N}_2 X^2$ where
$X$ is a Gaussian variable with $\langle X \rangle =0$ and $\langle X^2 \rangle = 1$. In this limit it can be
readily seen that the optimum result is $F_Q \simeq  30 \bar{N}^4$  that holds for $\bar{N}_2 \simeq 0.71 \bar{N}$,
so that for every $\phi$
\begin{equation}
\label{Df3}
\Delta^2 \phi \simeq \frac{1}{30 \bar{N}^4} .
\end{equation}

Comparing Eqs. (\ref{sM}) and (\ref{Df3}) we can appreciate that the resolution provided by the complete
number statistics $p_n$ outperforms the simple estimation after the measurement of $M$. This can be
regarded as a generalization to nonlinear interferometry of the result for the linear case in Ref. \cite{sv2}.
It is also worth noting that both in the linear and nonlinear schemes optimum results are obtained for the same
probe states. The only difference is the amount of squeezing, 50 \% in the linear case versus 70 \%
in the nonlinear one.  This is to say that nonlinearity offers resolution improvement without any drawback
in the probe preparation.

\subsection{Product of number states}
The above cases refer to realistic probe states. This can be compared with more ideal probes, such as
the product of number states $| \psi \rangle = | n_1 \rangle | n_2 \rangle$ with $n_1 + n_2 = \bar{N}$.
Starting with the simple estimator in Eq. (\ref{Df}) and using the complete exact expression Eq. (\ref{MSC})
we get
\begin{equation}
\langle M \rangle = ( n_1 - n_2 ) C_1,
\end{equation}
and
\begin{equation}
 \langle M^2 \rangle =  ( n_1 - n_2 ) C_1^2+  (2 n_1  n_2 +  n_1 + n_2  )  S^2_1,
\end{equation}
where $S_1 = \sin [2 \phi ( \bar{N} -1) + \phi_0]$,  $C_1 = \cos [2 \phi (\bar{N}-1)+ \phi_0]$, so that
for any $\phi$ and $\phi_0$
\begin{equation}
\label{Df4}
\Delta^2 \phi = \frac{2 n_1  n_2 +  n_1 + n_2 }{4( n_1 + n_2 -1)^2 (n_1  - n_2)^2  } .
\end{equation}

The minimum uncertainty holds either when $n_1=0$ or $n_2 =0$, so that for large $\bar{N}$ the
uncertainty scales as
\begin{equation}
\label{DfSU2}
\Delta^2 \phi \simeq \frac{1}{4 \bar{N}^3}.
\end{equation}
This input probe is an SU(2) coherent state, which are the projection on fixed total number
of the standard coherent states \cite{AD}. Accordingly the resolution is the same reached
with coherent probes in Eq. (\ref{Dfv}). However, note that in this case the result holds for all
$\phi$ because here there is no dispersive effect caused by the fluctuations of $N$. Beyond
this minimum, the uncertainty (\ref{Df4}) grows without limit as $n_1$ approaches $n_2$.
Deep down this holds because for $n_1=n_2$ we get that $\langle M \rangle$ no
longer depends on the signal $\phi$.

Regarding the more advanced estimation provided by the complete statistics $p_n$, we get the
opposite conclusions. After the Fisher information $F = F_Q = 4 (n_1 + n_2 )^2 ( 2 n_1  n_2
+  n_1 + n_2 )$  the minimum uncertainty holds for $n_1$ as close as possible to $n_2$, this
is $n_1= n_2 = \bar{N}/2$ for even $\bar{N}$, which is the well-known case of twin photon
states \cite{HB}. This leads to $F_Q \simeq 2 \bar{N}^4$ and
\begin{equation}
\label{Df5}
\Delta^2 \phi \simeq \frac{1}{2 \bar{N}^4},
\end{equation}
On the other hand, the minimum $F_Q$ holds for the SU(2) coherent states $n_1n_2 =0$ with
$F_Q \simeq 4 \bar{N}^3$ in agreement with Eq. (\ref{DfSU2}).

In this particular case, it is possible to reach the optimum resolution (\ref{Df5}) via the simple estimation
procedure  considering that the measured observable is $M^2$ instead of $M$. To simplify the calculation
we consider from the start the probe with $n_1= n_2 = \bar{N}/2 \gg 1$ for even $\bar{N}$. In such a case
\begin{equation}
\langle M^2 \rangle \simeq  \frac{\bar{N}^2}{2} S_1^2,
\end{equation}
and
\begin{equation}
\Delta^2 M^2 \simeq S_1^2  \left ( \frac{\bar{N}^4}{8} S_1^2  + 2 \bar{N}^2 C_1^2 \right )  .
\end{equation}
Considering $\phi_0 = 0$ and $\phi \rightarrow 0$ we get finally
\begin{equation}
\label{Df6}
\Delta^2 \phi \simeq \frac{1}{2 \bar{N}^4}  ,
\end{equation}
reaching the minimum value predicted by the Cram\'er-Rao bound in Eq. (\ref{Df5}). Incidentally
with the above computations it can be easily checked that the product of twin number states tend
to be minimum uncertainty states of the pair $G$, $M^2$ for $\phi_0 = 0$ as $\phi \rightarrow 0$, this is
$\Delta G \Delta M^2 \simeq | \langle [ G, M^2 ]\rangle  |/2 $.

\section{Conclusions}

We have examined the performance of nonlinear gyroscopes in the quantum regime that highlight some relevant
features for quantum metrology. This schemes does not imply length variations, so that this can be a built-in solid
detector where all the potential advantages of nonlinearity can be used without  the drawbacks caused if
lengths were allowed to vary. Then the interferometer can be made of optical fibers where length and field
confinement can be much improve the nonlinear effects.

A key result is that the optimum resolution can be approached by feasible coherent-squeezed inputs. This is
a translation to nonlinear detection of the same result already proved for linear schemes. We find remarkable
that in this non-linear scheme where the linear and non-linear propagations commute the
benefits of nonlinear schemes can be obtained for the same probe states of the linear interferometry.
However this may not be the case if the corresponding generators do not commute.

\section*{Acknowledgements}
We thank Prof.  L. Pezz\'e for helpful comments.
We acknowledge financial support from Spanish Ministerio
de Econom\'ia y Competitividad Projects No. FIS2012-33152
and No. FIS2012-35583 and from the Comunidad Aut\'onoma
de Madrid research consortium QUITEMAD+ Grant No.
S2013/ICE-2801.

\appendix

\smallskip

\section{Noise analysis}
For completeness let us address a simple but meaningful analysis of the effect  of unavoidable sources
of noise, such as finite quantum efficiency, thermalization and phase randomization. To be more specific,
we focus on the cases of pure coherent and coherent-squeezed probes since they provide the most
interesting, meaningful and practical of the situations studied above. Several sources of noise can be
readily accounted for by the replacement in the detection operator $M$ in Eq. (\ref{M})
 \begin{equation}
 \label{tf}
a^\dagger_1 a_2 \rightarrow e^{i \varphi} \left (\sqrt{\eta} a^\dagger_1 +  \sqrt{1- \eta} \,
b^\dagger_1 \right ) \left (  \sqrt{\eta} a_2 + \sqrt{1- \eta} \, b_2 \right ) ,
 \end{equation}
where $\varphi$ is a random phase that we will assume to be Gaussian distributed with zero mean
and variance $\sigma^2$, $\eta$ is the quantum efficiency of the detectors, and $b_j$ are
uncorrelated field modes in thermal states  with  $ \langle b_1 \rangle = \langle b_2  \rangle =
\langle b^\dagger_1 b_2 \rangle =0$, and  $(1-\eta) \langle b^\dagger_1 b_1 \rangle =
(1-\eta)  \langle b^\dagger_2 b_2 \rangle = N_t /2$ with $N_t  \ll N$, respecting  that quantum
efficiency and thermalization are independent sources of uncertainty.

In such a case for probes in the product of coherent and squeezed states we have that in the absence of
signal
 \begin{widetext}
 \begin{equation}
 \Delta^2 M = \langle M^2 \rangle = \eta^2 \left ( \alpha^2 \langle a_2^2 \rangle e^{- 2 \sigma^2}  +
 \alpha^2 \langle a_2^{\dagger 2} \rangle e^{- 2 \sigma^2}  +  \alpha^2 \langle a^\dagger_2 a_2 \rangle
+  \langle a^\dagger_2 a_2 \rangle \right ) + \eta \left (  \alpha^2 + \langle a^\dagger_2 a_2 \rangle \right )
\left ( N_t +1 -\eta \right ) .
\end{equation}
\end{widetext}
where $\alpha$ is the coherent amplitude assumed real. Taking into account that in our case $\langle
a_2^2 \rangle = - \cosh r \sinh r$, $\langle a^\dagger_2 a_2 \rangle = \sinh^2 r$, assuming $r \gg1$, and
considering  the optimum case where the number of photons in the squeezed state is  $\langle a^\dagger_2
a_2 \rangle = \bar{N}_2 \simeq  \sqrt{\bar{N}}/2 \ll \bar{N}$,  we get
 \begin{equation}
 \Delta^2 M = \eta^2 \sqrt{\bar{N}}+ \eta^2 \bar{N}^{3/2} \epsilon + \eta \bar{N} \left ( N_t + 1 -\eta \right ) .
 \end{equation}
 where $\epsilon = 1-\exp(-2 \sigma^2 ) \simeq 2 \sigma^2$,  while for the denominator in Eq. (\ref{Df})
 we have just $\left | \partial \langle M \rangle / \partial \phi \right | \simeq 2 \eta^2 N^2$. Therefore the final
 form for the signal uncertainty becomes
 \begin{equation}
 \Delta^2 \phi \simeq  \frac{1}{4 \eta^2 \bar{N}^{7/2}} + \frac{N_t+1-\eta}{4 \eta^3 \bar{N}^3}  +
 \frac{\epsilon}{4 \eta^2 \bar{N}^{5/2}}  .
 \end{equation}
The most potentially harmful is the last term caused by the fluctuations of the relative phase $\varphi$. This
is natural because variations of $\varphi$ cause that the coherent field couples with the anti-squeezed
component  of the vacuum mode $a_2$, so the planned squeezing reduction becomes actually noise
amplification. This effect would spoil the improvement caused by the squeezing vacuum unless $\epsilon
\simeq 2\sigma^2 \leq 1/\bar{N}$. The other two terms in $\Delta^2 \phi$  are due to finite quantum efficiency
and thermal photons, and would tend to phase uncertainty similar to the case of pure coherent probes unless
$N_t + 1 - \eta \leq 1/\sqrt{\bar{N}}$.

 On the other hand, for the case of pure coherent probes we get  that the uncertainty preserves the scaling
 \begin{equation}
 \Delta^2 \phi \simeq   \frac{N_t+1}{4 \eta^3 \bar{N}^3} .
 \end{equation}
in accordance with the robustness of coherent light.

\bigskip

\section{Linear and nonlinear phase shifts}

A relevant characteristics of the propagation in optically nonlinear media is that the field experiences
always both linear and nonlinear effects. This raises a very interesting question regarding which
transformation will encode optimally the signal, and whether its performance would be affected by
the presence of the other one. The proper arena to examine this issue is a multi-parameter estimation
procedure \cite{Ch}. More specifically, if the signal-dependent transformation is
\begin{equation}
U_\Omega = e^{-i \phi_L G_L - i \phi_{NL} G_{NL} } ,
\end{equation}
lower bounds for the estimation of $\phi_L$ and $\phi_{NL}$ can be obtained in terms of the
quantum Fisher information matrix as
\begin{widetext}
\begin{equation}
F = 4 \pmatrix{ \Delta^2 G_L & \mathrm{Re} \langle G_L G_{NL} \rangle - \langle G_L \rangle
\langle G_{NL} \rangle \cr \mathrm{Re} \langle G_L G_{NL} \rangle - \langle G_L \rangle \langle G_{NL}
\rangle  & \Delta^2 G_{NL} } ,
\end{equation}
\end{widetext}
as
\begin{equation}
\Delta^2 \phi_L \geq \left ( F^{-1} \right )_{1,1} , \quad \Delta^2 \phi_{NL} \geq \left ( F^{-1} \right )_{2,2} .
\end{equation}
In our case we have
\begin{equation}
G_L = i \left ( a^\dagger_2 a_1 - a^\dagger_1 a_2 \right ) , \quad G_{NL} = N i \left ( a^\dagger_2 a_1 -
a^\dagger_1 a_2 \right )  ,
\end{equation}
with $[G_L,G_{NL} ]=0$. Since we are dealing with quantum Fisher information we
may follow the same approximations leading to Eq. (\ref{Df3}).

After an straightforward calculation we finally get that for fixed mean total photon number $\bar{N}$
the optimum $\Delta^2 \phi_L$ holds for $\bar{N}_2 \simeq 0.6 \bar{N}$ leading to $\Delta^2 \phi_L
\simeq 0.3/\bar{N}^2$ which is rather close to the Heisenberg limit $0.25/\bar{N}^2$.

On the other hand the optimum $\Delta^2 \phi_{NL}$ holds for $\bar{N}_2 \simeq 0.8 \bar{N}$ leading
to $\Delta^2 \phi_{NL} \simeq 0.04/\bar{N}^4$ which is also rather close to Eq. (\ref{Df3}). Therefore, we
may say that the multi-parameter protocol reproduces expected results for the estimation of linear and
nonlinear phase shifts which seem to be not affected by the presence of the other part.

We have carried out the same analysis for the case when there is no squeezing so the mode $a_2$
is in the plain vacuum state leading to, in the limit $\bar{N} \gg 1$, to  $\Delta^2 \phi_L \simeq 1/4$
and  $\Delta^2 \phi_{NL} \simeq 0.25 /\bar{N}^2$. So in comparison with the squeezed case the
conclusion would be the opposite in the sense that there would be a clear perturbation between
linear and nonlinear estimation processes, as already encountered in an slightly different single-mode
nonlinear detector scheme  \cite{Ch}.

\end{document}